\documentclass{article}

\usepackage{PRIMEarxiv}

\usepackage[utf8]{inputenc} 
\usepackage[T1]{fontenc}    
\usepackage{hyperref}       
\usepackage{url}            
\usepackage{nicefrac}       
\usepackage{microtype}      
\usepackage{lipsum}
\usepackage{fancyhdr}       
\usepackage{graphicx}       
\graphicspath{{media/}}     

\usepackage{booktabs}
\usepackage{balance}
\usepackage{amsfonts}
\usepackage{amsmath}
\usepackage{amsbsy}
\usepackage{cases}
\usepackage{bm}
\usepackage{algorithm}
\usepackage{multirow}
\usepackage{booktabs}
\usepackage{cases}

\usepackage{svg}
\usepackage{float}
\usepackage{algpseudocode}
\algrenewcommand\algorithmicrequire{\textbf{Input:}}
\algrenewcommand\algorithmicensure{\textbf{Output:}}
\usepackage{cleveref}
\usepackage{caption}
\usepackage{subcaption}
\usepackage{braket}
\usepackage{mathtools}
\DeclarePairedDelimiterX{\infdivx}[2]{(}{)}{%
  #1\;\delimsize\|\;#2%
}

\newcommand{\paragraStartHighlight}[1]{\noindent\textbf{#1}}

\DeclareMathOperator{\concat}{\text{\textbf{CONCAT}}}

\DeclareMathOperator{\mlp}{\text{\textbf{MLP}}}
\DeclareMathOperator{\linear}{\text{\textbf{LINEAR}}}

\DeclareMathOperator{\sigmoid}{\text{Sigmoid}}
\DeclareMathOperator{\enc}{\text{\textbf{ENC}}}
\DeclareMathOperator{\dec}{\text{\textbf{DEC}}}
\DeclareMathOperator{\agg}{\text{\textbf{AGGREGATE}}}
\DeclareMathOperator{\combine}{\text{\textbf{COMBINE}}}
\DeclareMathOperator{\fr}{\text{\textbf{FR}}}

\newcommand{\proposedModel}{JMMFR-MC} 

\newtheorem{problem}{Problem}
\crefrangelabelformat{problem}{(#3#1#4--#5\crefstripprefix{#1}{#2}#6)}

\newtheorem{definition}{Definition}

\title{Remote Work Optimization with Robust Multi-channel Graph Neural Networks} 

\author{
  Qinyi Zhu \\
  University of California, Berkeley \\
  \texttt{qinyi\_zhu@berkeley.edu} \\
   \And
  Liang Wu, Qi Guo, Liangjie Hong \\
  LinkedIn \\
  \texttt{\{liawu, qiguo, liahong\}@linkedin.com} \\ 
}

\begin{document}
\maketitle
\begin{abstract}
The spread of COVID-19 leads to the global shutdown of many corporate offices, and encourages companies to open more opportunities that allow employees to work from a remote location. As the workplace type expands from onsite offices to remote areas, an emerging challenge for an online hiring marketplace is how these remote opportunities and user intentions to work remotely can be modeled and matched without prior information. Despite the unprecedented amount of remote jobs posted amid COVID-19, there is no existing approach that can be directly applied.

Introducing a brand new workplace type naturally leads to the cold-start problem, which is particularly more common for less active job seekers. It is challenging, if not impossible, to onboard a new workplace type for any predictive model if existing information sources can provide little information related to a new category of jobs, including data from resumes and job descriptions. Hence, in this work, we aim to propose a principled approach that jointly models the remoteness of job seekers and job opportunities with limited information, which also suffices the needs of web-scale applications. Existing research on the emerging type of remote workplace mainly focuses on qualitative studies, and classic predictive modeling approaches are inapplicable considering the problem of cold-start and information scarcity. We precisely try to close this gap with a novel graph neural architecture. Extensive experiments on large-scale data from real-world applications have been conducted to validate the superiority of the proposed approach over competitive baselines. The improvement may translate to more rapid onboarding of the new workplace type that can benefit job seekers who are interested in working remotely.
\end{abstract}

\keywords{Hiring marketplace \and Experimentation \and Remote work}

\section{Introduction}
The COVID-19 pandemic is quickly changing how every organization is attracting and recruiting employees on their virtual teams, making remote work the new normal. The traditional job recommender systems~\cite{Shibbir2016,Bansal2017,Shalaby2017,Ruijt2021,Upadhyay2021} were designed for onsite opportunities instead of remote ones, which are already showing signs of not being able to scale and meet the challenges of the new post-COVID-19 world since it fails to consider the remote workplace. A new workplace model is needed to fill the gap in the existing job recommender systems by leveraging the remote workplace type, which enhances the matching between remote job seekers and vacancies. The content-based and graph-based models~\cite{Kipf2016,Hamilton2017,Velickovic2017,Tang2020} can thus be adopted to handle this problem. Content information, such as the content analysis of resumes and job descriptions, and graph structures, such as member-job interactions, can be utilized in the prediction of remoteness. 

A critical limitation of the remote workplace prediction is the lack of learnable features. Many studies begin to pay attention to the remote workers and positions~\cite{Brynjolfsson2020,Dey2020,Dingel2020,Krantz-Kent2019,Mongey2020}, while they only conduct surveys and discuss the necessity to consider the remote work without further experiments. Recent work estimates that $31\%$ workers had switched to working at home during pandemic~\cite{Brynjolfsson2020} and around $37\%$ jobs can be performed entirely at home~\cite{Dingel2020}. All the evidence suggests the importance of introducing the new workplace type prediction to provide more personalized career recommendations, while it has not yet been executed. One main obstacle is the limitation of high-quality member and job information, which plays a crucial role in the successful predictions. In order to answer questions like "do workers prefer to work remotely?" and "can jobs be performed remotely?", the availability of high-quality and reliable information sources is of great importance, such as member skills, job titles, industries and prior job applications. The workplace predictions are particularly hard when these meaningful data are missing. It's almost impossible to predict the remoteness preference for members and jobs based on low-quality information, even for best-performing models.

In this paper, we focus on learning the remoteness preference with both the content-based and graph-based information, specifically on the real-world datasets with numerous missing features. Our motivations are mainly inspired by the following gaps:

\paragraStartHighlight{Missing Feature/Cold Start} problem widely exists in the real-world datasets~\cite{Yuan2016,Friedjungov2019,Jiang2020,Taguchi2021,Rossi2021,Chen2022,Leitch2022}, while conventional networks normally assume a complete set of trainable features, which is hard to achieve in the online recruitment markets. For example, one less active member may refuse to provide any information, leaving an empty node feature. For the behavioural data, it's also possible that there is no member-job interaction data for some members. Imputation treatments for missing values have been widely investigated~\cite{Zhang2013,Zhu2015,Thirukumaran2016},
and plenty of graph-based methods have utilized the graph structures to reconstruct missing data~\cite{Rossi2021,Jiang2020,Taguchi2021,Chen2022}, but these methods are not directly designed for the hiring marketplace, since the workplace type problem is inherently cold-start~\cite{Guo2017,Lian2017}: less active members/jobs don't have any textual and behavioral feedback related to remote work, making the model learning difficult.

\paragraStartHighlight{Scalability} is crucial in the online systems. Different from most existing work~\cite{Rossi2021}, our approach doesn't involve the adjacency matrices, which are extremely expensive for a large number of nodes. 
With the vast amount of incoming interactions every day, building and maintaining a graph-based system is not trivial. 
Besides that, multi-channel learning is used to further decrease the feature dimension.
Job models usually use the vector representations of categorical variables, such as one-hot encoding. This strategy breaks down when the number of categories grows, as it creates high-dimensional feature vectors. Hence it's important to scale down the size of network while maintaining its ability~\cite{Zhou2019, Cao2019,Hu2021}, to adapt to the limited memory on the online platforms. 

\paragraStartHighlight{Low-Degree Nodes} are much more common in the job application dataset than the other social networks. Graph structures can be captured using graph models~\cite{Kipf2016,Hamilton2017,Velickovic2017,Tang2020}, while the usage of graph models is less explored in the job-member systems~\cite{Shalaby2017}. Graph Neural Networks (GNNs) have been actively used in both link prediction and node classification problems, and are widely proven to have promising and remarkable accuracy. However, GNNs normally assume a rich neighborhood with a large number of neighbors, which is hard to achieve in the job application dataset~\cite{Yuan2016,Guo2017,Lian2017}. The real-world job application dataset often lacks behavioral features, especially for new users, which call for a specific design to resolve the issue.

In this work, we propose a job-member embedding model with feature restoration through multi-channel learning (JMMFR-MC) to close the gaps above. Our
model aims at filling the gap in existing online recruitment systems by predicting the remote workplace types and tackling the missing feature problem through multi-channel learning in one unified salable graph-based architecture. At a high level, our system consists of four main steps. First, we build a bipartite graph of members and jobs as nodes with member-job interactions as edges. Second, multi-channel learning is built to capture various types of characteristics. Third, a feature restoration operator is defined to recover the missing features. Finally, the workplace type prediction and missing feature problems are learned jointly. Our main contributions are summarized as:
\begin{itemize}
    \item We define a bipartite member-job graph to enable large-scale learning on the real-world dataset.
    \item We propose a feature restoration algorithm to recover the missing features using textual content and graph structure.
    \item We formalize the training task as a multi-object learning problem to learn the task-specific features.
    \item We experimentally validate that our approach outperforms the state-of-the-art models in real-world scenarios and provide an in-depth analysis of the robustness.
\end{itemize}
The rest of this paper is organized as follows. In Section~\ref{sec:problem-definition}, we provide the definition for the member-job graph and corresponding problem statements. In Section~\ref{sec:proposed-approach}, we recap the state-of-art graph models and introduce our proposed approaches. In Section~\ref{sec:experiment}, we experimentally validate the performance of our proposed approaches and analyze the model behavior. We discuss the related work in Section~\ref{sec:related-work}, and the conclusion and future work in Section~\ref{sec:con}.

\section{Problem Statement}\label{sec:problem-definition}

\begin{table}[h!]
  \centering
  \caption{Notations and Description}
  \label{tab:notation}
  \begin{tabular}{p{2cm} p{5cm}}
    \toprule
    Notation & Description \\
    \midrule
    $\mathcal{G}$ & The member-job bipartite undirected graph that captures job application\\
    $\mathcal{U}$ & Set of members in the graph\\
    $\mathcal{V}$ & Set of jobs in the graph\\
    $\mathcal{E}$ & Set of edges between members and jobs\\
    $\mathbf{X}$ & Set of node features\\
    $\mathbf{Y}$ & Set of node-level labels\\
    \bottomrule
  \end{tabular}
\end{table}

We use the following formulation,
\begin{equation}
    \mathcal{G}=(\mathcal{U},\ \mathcal{V},\ \mathcal{E}), \ \mathbf{X},\ \mathbf{Y},
\end{equation}
to denote the whole bipartite undirected graph, node-level features and labels, where $u_i\in\mathcal{U}$ ($v_i\in\mathcal{V}$) represents one member (job) in the graph, $e_{ij}=e_{u_iv_j}\in\mathcal{E}\subseteq\mathcal{U}\times\mathcal{V}$ represents the undirected edge between one member $u_i$ and one job $v_j$, vector $\mathbf{x}_i\in\mathbf{X}$ and scalar $y_i\in\mathbf{Y}$ denote the node-level features and labels respectively for both the members and jobs. Note that members and jobs share the same feature space. For some members and jobs, their node features are empty vectors. Given one member $u_i$, the subgraph 
\begin{equation}
    \mathcal{G}_i=(u_i,\ \mathcal{N}(u_i),\ (u_i\times\mathcal{N}(u_i))\cap\mathcal{E}),\ \mathbf{X},\ y_i,
\end{equation}
includes the member $u_i$ and its corresponding neighborhood $\mathcal{N}(u_i)$. The subgraph of jobs are similar. Given those terminologies, we are able to define our problems:

\begin{problem}[Remoteness Prediction] \label{p1}
Given an undirected bipartite graph structure $\mathcal{G}$, node features $\mathbf{X}$, neural network $f_{\theta}(\cdot)$ with \textbf{learnable} parameters $\theta$, we use the nodes' neighborhood $\mathcal{G}_i$ to predict their remoteness preference $f_{\theta}(\mathbf{X},\ \mathcal{G}_i)\in [0,1]$. Our goal is to find the best model $f_{\theta}(\cdot)$ that optimizes the remoteness accuracy.
\end{problem}

In Problem \ref{p1}, the conventional model $f_{\theta}(\cdot)$ normally assumes a complete set of non-empty node/edge features, which is hard to achieve in the real-world data. The real-world job seeker data often lack the high-quality features. For example, one user may prefer not to provide any self-information, leaving an empty node feature. For traditional neural network, it's impossible to train and learn with an empty feature. However, the existence of edges in the graph gives it possibility to learn this user's behaviour from its interactions. Moreover, it gives us the possibility to restore one node's features from its own neighborhood. In Problem \ref{p2}, we propose a Feature Restoration operation to restore the source node features from the source nodes' neighborhood and use the restored information to compensate for the missing features.

\begin{problem}[Feature Restoration] \label{p2}
Given an undirected bipartite graph structure $\mathcal{G}$, feature restoration operator $g_w(\cdot)$ with \textbf{learnable} edge-wise weights $w$, our task is to find the best weights that are able to recover nodes' missing features $g_{w}(\mathbf{X},\ \mathcal{G}_i)$ using the nodes' neighborhood information $\mathcal{G}_i$.
\end{problem}

Using the feature restoration in Problem \ref{p2}, we are able to obtain the ``\emph{task-irrelevant}'' features. In practice, feature restoration is often treated as a preprocessing step and learned in advance of the main task. Here, in order to restore the \emph{task-relevant} features and extract the segments more related to our remoteness prediction task, we train the feature restoration operator and the remoteness prediction task jointly as a multi-object learning task, as in Problem \ref{p3}.

\begin{problem}[Multi-Object Learning] \label{p3}
Given an undirected bipartite graph structure $\mathcal{G}$, neural network $f_{\theta}(\cdot)$ with \textbf{learnable} parameters $\theta$, feature restoration $g_w(\cdot)$ with \textbf{learnable} weights $w$, we first restore the source nodes' missing features, then use the restored node features and the original node features (if exist) to predict the nodes' remoteness preference. Our goal is to find the best feature restoration operator and neural model via the same training process.
\end{problem}

\section{Proposed Approach: JMMFR-MC --- \underline{J}ob-\underline{M}ember embedding \underline{M}odel with \underline{F}eature \underline{R}estoration through \underline{M}ulti-\underline{C}hannel learning}\label{sec:proposed-approach}
Multiple aspects affect why one member applied for one job. If one member applied for one job, it may imply that this member and job share some similarities in various aspects, such as this member and job are both in artificial intelligence industry; this member has the skills this job is asking for.
Moreover, different types of aspects also affect the members' and jobs' willingness to take and offer remote opportunities to some extent. For example, members and jobs from tech industries may prefer the work-from-home policy, while most skill information may not directly affect the remoteness preference as much as the industry information, such as art design skill and Microsoft Office skill. Therefore, given a specific aspect, we would like to learn the aspect-specific similarities between members and jobs, then use these similarities to restore the missing features and predict the remoteness intention. The main process can be constructed as: 1) one member applied for one software engineer job and this job requires the Microsoft Office skills, so we predict this member has a probability $p_1$ to be a software engineer and have a probability $p_2$ to own Microsoft Office skills. These probabilities soften the one-hot encoding features of this member from $[0,1,0,\cdots]$ into the restored feature $[0,1,0,\cdots,p_1,\cdots,p_2,\cdots]$; 2) given the information that this member maybe a software engineer with Microsoft Office skills, we predict the remoteness possibility of this member using the restored feature including $p_1,p_2$.

In order to generate the probabilities $p_1,p_2$ and then use them to predict the remoteness, aspect-specific multi-channel learning is used. One node's possible skills only depend on its neighbors' possible skills, and aren't affected by any other aspects, such as industries. There are two main reasons about this approach: 1) The dimension of the skill subspace is hundreds of times larger than the dimension of the industry subspace, and both the skill and industry features are very sparse, sometimes even empty. Taking the concatenation of the skill and industry feature brings the difficulty in finding the global minimum. 2) The node features include a combination of various aspects. These aspects affect the member-job relation differently and evolve between nodes up to different scale. For example, skill features are one-hot vectors, while title features are embeddings. Besides that, the variety of skills is much larger than the titles and industries.

In this Section, we first introduce the workplace type prediction problem and content-based approaches in Section~\ref{sec:workplaceType}, then recap the basic knowledge in Graph Neural Networks, introducing the main difference between graph models and our feature restoration treatment in Section~\ref{sec:graph-network}. Then we introduce the two main techniques in our work -- multi-channel learning in Section~\ref{sec:multi-channel} and feature restoration operator in Section~\ref{sec:feature-restoration}. Finally, we summarize our work and give the whole learning workflow in Section \ref{sec:forward-backward}.

\subsection{Content-Based Approaches}\label{sec:workplaceType}
Due to the heterogeneity of nodes~\cite{Wang2019,Hu2020}, different types of nodes have different feature distribution. Therefore, for each type of nodes (e.g., member nodes and job nodes), we design the type-specific transformation matrix to project the features of different nodes types into the same
feature space. Unlike heterogeneous graph models~\cite{Hamilton2018,Wang2019,Hu2020}, the type-specific transformation matrix is based on node types rather than edge types. Therefore, for members and jobs, we use the different linear transformation matrix to project the node features into the low-dimensional space. The linear projection operator $\linear^{\text{\emph{node}},\ N_{\text{in}}}_{N_{\text{out}}}(\cdot)$ can be formulated as follows:
\begin{equation*}
\begin{split}
\mathbf{x}_i&\leftarrow\linear^{\text{\emph{mem}},\ N_{\text{in}}}_{N_{\text{out}}}(\mathbf{x}_i),\ \ \ \text{for}\ u_i\in\mathcal{U},\\
\mathbf{x}_j&\leftarrow\linear^{\text{\emph{job}},\ N_{\text{in}}}_{N_{\text{out}}}(\mathbf{x}_j),\ \ \ \ \ \text{for}\ v_j\in\mathcal{V},
\end{split}
\end{equation*}
where $N_{\text{in}}$ and $N_{\text{out}}$ are the dimensions of the original and projected node features respectively, $\linear^{N_{\text{in}}}_{N_{\text{out}}}(\cdot):\mathbb{R}^{N_{\text{in}}}\to\mathbb{R}^{N_{\text{out}}}$ is a $\mathbb{R}^{N_{\text{out}}}$-by-$\mathbb{R}^{N_{\text{in}}}$ matrix with $N_{\text{in}}\leq N_{\text{out}}$. Note that the input and output dimensions vary among channels.

After feature transformation, we are able to leverage the content-based models to learn the remote workplace types.
Given the node feature $\mathbf{x}_i\in\mathbf{X}$, the content-based models follow the formula:
\begin{equation*}
\begin{split}
\mathbf{z}_i&=\mlp(\mathbf{x}_i),\\
\bar{y}_i&=\dec(\mathbf{z}_i),
\end{split}
\end{equation*}
where $\mathbf{x}_i$, $\mathbf{z}_i$ and $\bar{y}_i$ are the node feature, hidden state and workplace type prediction of node $v_i$ respectively, $\mlp(\cdot)$ is the fully-connected layers, $\dec(\cdot)$ is the single-layer linear decoder and more sophisticated design can be used based on the research needs. Most existing works on job systems focused on the content-based models. However, the power of content-based models is limited and constrained by the semantic analysis on the textual content.

\subsection{Graph-Based Approaches}\label{sec:graph-network}
\begin{definition}[Embedding Space]
An embedding space $\mathcal{Z}\subseteq\mathbb{R}^{d_\mathcal{Z}}$ is a vector space onto which we project nodes $\mathcal{V}$. Embedding space is normally equipped with a metric $S_\mathcal{Z}(\cdot,\cdot)$ such that $s_{ij}=S_\mathcal{Z}(\mathbf{z}_i, \mathbf{z}_j)$ denotes a similarity measure between two embeddings.
\end{definition}

Graph Neural Networks normally follow an encoder-decoder framework~\cite{Kipf2016,Hamilton2017,Velickovic2017,Zhu2021}. We let $\mathcal{G}_i$ denote the neighborhood\footnote{A $L$-hop neighborhood$\mathcal{N}_i^L$ considering all the nodes and edges within a radius of $L$-edges w.r.t $i$ is a widely used definition.} around node $u_i$, which comprises nodes, edges, node features or possible edge features if exist. Encoders extract the local neighborhood information $\mathcal{G}_i$ to get a more contextual node representation $z_i$ in the embedding space.
In principle, the similarity between two connected nodes in the embedding space are higher than the similarity between two unconnected nodes. 
Then decoders convert the node representation into desired output and the output formula is designed specifically for various tasks. For node classification task, decoders map the node representation $z_i$ into the node classes. For link prediction task, decoders measure the similarity $s_{ij}$ between two node representations $z_i$ and $z_j$. Conceptually an encoder $\enc: \mathcal{U}\cup\mathcal{V} \rightarrow \mathbb{R}^{d_\mathbf{Z}}$ projects nodes to an embedding space whereas the actual model maps neighborhood $\mathcal{G}_i$  to an embedding $\mathbf{z}_i$. The concrete choices of decoders vary in different applications. For example, in our node-level supervised learning with labels $y_i\in\mathbf{Y}$, our goal is to optimize the parameters in the encoder and decoder:  $$\min_{\enc, \dec} \mathbb{E}_{\mathcal{U},\mathcal{V}}[\mathcal{L}(\mathbf{Y},\ \dec(\enc(\mathbf{X},\ \mathcal{G})))].$$ 

Encoders contain the key idea of graph neural networks, including two important operators $\agg(\cdot)$ and $\combine(\cdot)$:
(a) $\agg(\cdot)$ operator aggregates the neighborhood information, including the target nodes' representations, edge information, using operations such as mean, sum, and LSTM, while more sophisticated pooling and normalization functions can be also designed. (b) $\combine(\cdot)$ operator combines the source node information with the aggregated neighborhood information using operations such as concatenation and addition, while more sophisticated neural layers can be also designed. Modern $L$-depth GNNs follow a neighborhood aggregation strategy, where we iteratively update the representation of a node by aggregating representations of its neighbors. Given a source node $u_i$, homogeneous encoders use the node neighborhood $\mathcal{N}(u_i)$, node and edge features to learn a node embedding $\mathbf{z}_i\in\mathbf{Z}$.
After $L$ iterations of aggregation, a node’s representation captures the structural
information within its $L$-hop network neighborhood. Formally, the $l$-th layer of a GNN is
\begin{equation}\label{eqn:enc-dec}
\begin{split}
\mathbf{a}_i^{(l)}&=\agg(\{\mathbf{z}_j^{(l-1)},\ v_j\in\mathcal{N}(u_i)\}),\\
\mathbf{z}_i^{(l)}&=\combine(\mathbf{z}_i^{(l-1)},\ \mathbf{a}_i^{(l)}),
\end{split}
\end{equation}
where $l\in[1,L]$, $\mathbf{z}_i^{(l)}$ is the node embedding of $u_i$ with the message of $l$-hop network neighbors,  $\mathbf{a}_i^{(l)}$ is the message aggregated from $l$-hop neighbors and we initialize $\mathbf{z}_i^{(0)}=\mathbf{x}_i\in\mathbf{X}$. 
In this work, attention mechanism or mean pool is used to aggregate the neighbor representations. Aggregated messages are further combined with self embeddings using fully connected layers.
We will drop superscripts and use $\mathbf{z}_i$ or $\mathbf{Z}$ to represent node embeddings in the following discussion.

Following the encoder-decoder framework, node neighborhood $\mathcal{G}_i$ is projected to a vector space $\mathbf{Z}$ which is then decoded for different tasks: a) \emph{Node-Level Task}, such as node classifications; b) \emph{Link-Level Task}, such as link predictions; c) \emph{Graph-Level Task}, such as predicting the property of an entire graph. In this work, we will mainly focus on the node-level task. Link-level or graph-level tasks are left for the future discussions.

For node-level tasks, the decoders and loss functions are written, 
\begin{equation*}
\begin{split}
\bar{y}_i&=\dec(\mathbf{z}_i),\\
\mathcal{L}&=\mathbb{E}_{\mathcal{U},\mathcal{V}}[\ell(\bar{y}_i,\ y_i)],
\end{split}
\end{equation*}
where $y_i\in \mathbf{Y}$, $\bar{y}_i$, $\mathbf{z}_i$ are the node label, node prediction and node embedding respectively. 

Using the graph structure, we are able to obtain the node embedding $\mathbf{z}_i$ even though the node feature is empty $\mathbf{x}_i=\mathbf{0}$. However, learning a good node embedding is challenging for graph with multiple aspects of features, even more challenging considering that there are missing features and the missing level varies among aspects. In order to tackle this problem, we would like to ask whether it is possible to restore the aspect-specific nodes' missing features using the graph structure? Even more, is it possible to restore the aspect-specific and task-relevant nodes' missing features using the graph structure? In this work, we adopt the idea of the multi-object learning and propose a model that are able to restore the task-relevant nodes' missing features in each aspect and argue for the importance of learning the task of our interest and the task-relevant missing features jointly.

\begin{figure}[h]
    \centering
    \includegraphics[width=\linewidth]{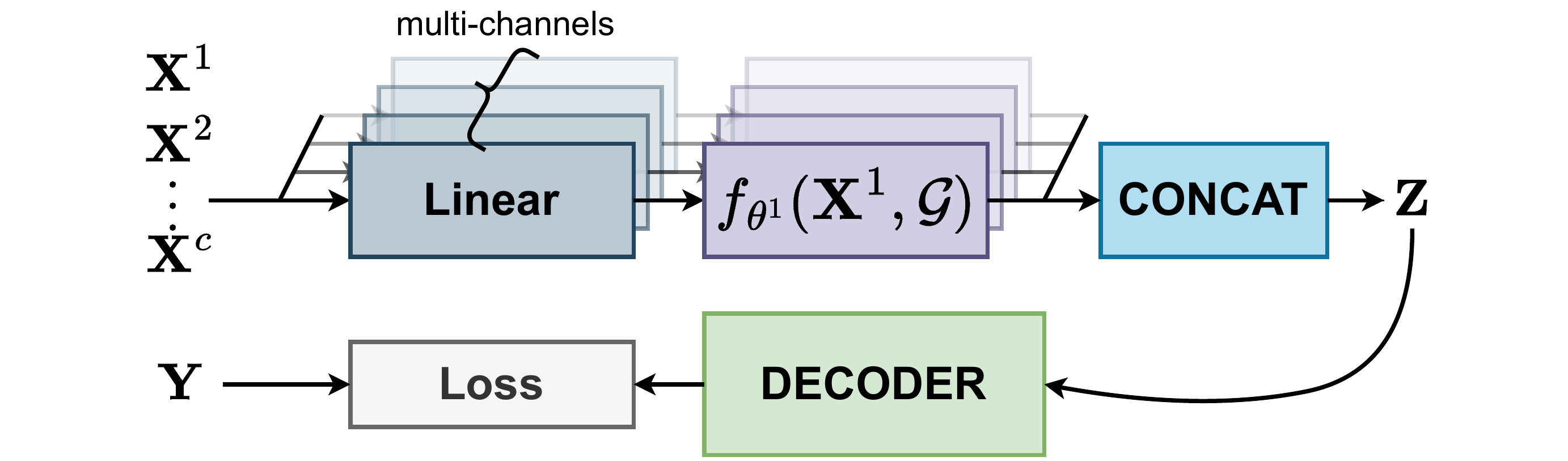}
    \caption{The diagram illustrating the multi-channel learning. Node feature space are split into various channels, such as node skills, industries and titles. Channels are independent from each other and fed into independent content-based or graph-based models to compute the embeddings. Finally, the embeddings from all the channels are concatenated together and passed to the decoder to predict the node remoteness.}
    \label{fig:multi-channel}
\end{figure}

\subsection{Multi-Channel Learning} \label{sec:multi-channel}
In the multi-channel learning, the node features are split into several aspects and the aspect-specific encoders are implemented, which means that aspects have independent parameters and don't interact with each other until the very last decoder step. One of the main challenges here is that \emph{``job seekers and recruiters do not always speak the same language''}~\cite{Schmitt2016}. Fortunately, in our dataset, the jobs and members always share the same feature space and this feature space is able to be split into a set of subspace, corresponding to different aspects, including skills, industries, titles, etc. This shared and divisible feature space benefits our multi-channel learning. Assume that we split the feature space into $N_c$ channels:
\begin{equation}
    \mathbf{X}=\{\mathbf{X}_1,\cdots,\mathbf{X}_c\},
\end{equation}
with $\dim(\mathbf{X})=\sum_{c=1}^{N_c}\dim(\mathbf{X}_c)$.

In our multi-channel learning, each channel is treated individually as opposed to combining all features by concatenation. Therefore, an aspect-specific node embedding $\mathbf{Z}^c=\{z^c_i\}_i$ is learned, 
\begin{equation}
\begin{split}
\mathbf{Z}^c&=f_{\theta^c}{(\mathbf{X}^c,\ \mathcal{G})},\ \ \ \text{for all nodes and channels } c,\\
\mathbf{Z}&=\concat{(\{\mathbf{Z}^c\}_c)},\\
\bar{\mathbf{Y}}&=\dec{(\mathbf{Z})},\\
\end{split}
\end{equation}
where $f_{\theta^c}(\cdot)$ denotes the content-based or graph-based approach. For content-based approach, graph structure $\mathcal{G}$ isn't used. For graph-based approach, graph structure is used and the $L$-hop graph-based approaches generally follow the encoder-decoder structure in Eq~\ref{eqn:enc-dec} with various $\agg(\cdot)$ and $\combine(\cdot)$ operators, including GCN~\cite{Kipf2016}, GraphSAGE~\cite{Hamilton2017}, GAT~\cite{Velickovic2017}. $\dec(\cdot)$ denotes the decoder predicting the workplace types using node embeddings via two fully-connected layers: 
\begin{equation}
\begin{split}
\mathbf{H}&=\sigma(\linear^{m}_d(\mathbf{Z}))\\
\bar{\mathbf{Y}}&=\sigmoid(\linear^{d}_1(\mathbf{H})),
\end{split}
\end{equation}
with hidden states $\mathbf{H}$, element-wise non-linear activation function $\sigma(\cdot)$, $\sigmoid(\cdot)$, linear transformation $\linear^{m}_d(\cdot):\ \mathbb{R}^{m}\to\mathbb{R}^{d}$, $\linear^{d}_1(\cdot):\ \mathbb{R}^{d}\to\mathbb{R}$.

Then this node embedding is used to predict the node remoteness and compared with the node-level labels:
\begin{equation}
\begin{split}
\min_{\theta^c}\mathcal{L}&=\mathbb{E}_{\mathcal{U},\mathcal{V}}[\ell(\bar{\mathbf{Y}},\ \mathbf{Y})],\\
&=\mathbb{E}_{\mathcal{U},\mathcal{V}}[\ell(\dec{(\concat{(\{f_{\theta^c}{(\mathbf{X}^c,\ \mathcal{G})}\}_c)})},\ \mathbf{Y})].
\end{split}
\end{equation}

\begin{figure}[h]
    \centering
    \includegraphics[width=0.9\linewidth]{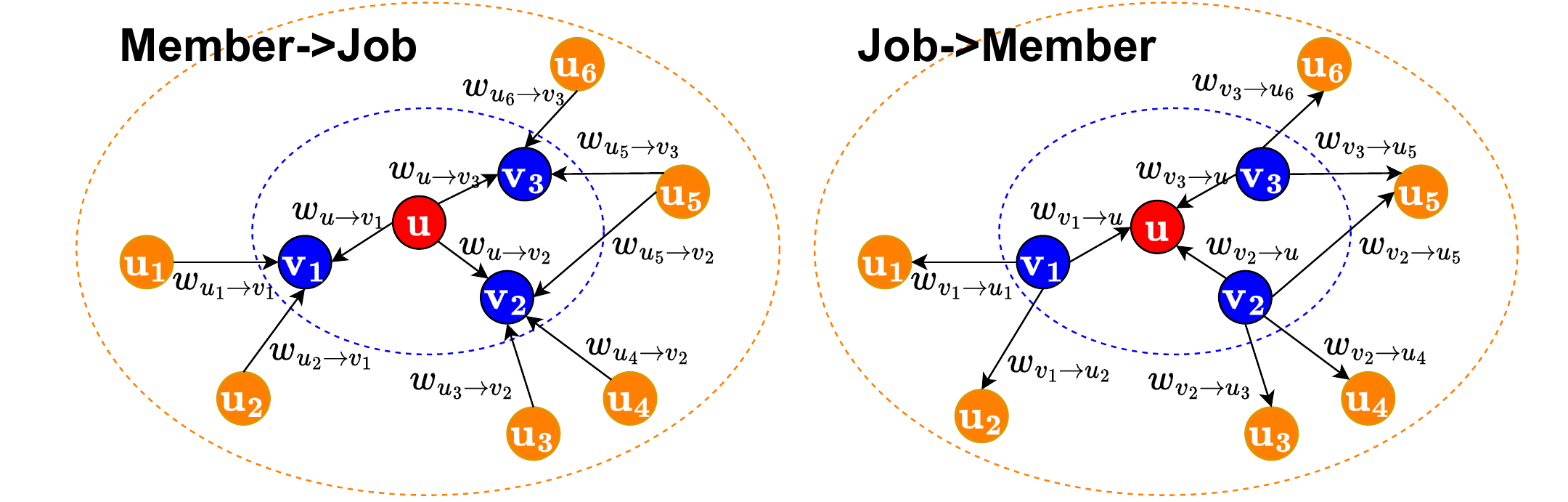}
    \caption{The diagram illustrating the Feature Restoration with edge-wise weights. The nodes' neighbor features are propagated over the edges to recover the node missing features. The original and restored node features are then used together to accomplish the remoteness prediction problem. Note that the weights $w_{u\to v}$ and $w_{v\to u}$ are different.}
    \label{fig:feature-propagation}
\end{figure}

\subsection{Feature Restoration} \label{sec:feature-restoration}
The lack of information limits our ability to predict the remoteness. GNNs normally assume a complete set of non-empty node features. However, in the real-world data, the lack of features and edges is common. In the member-job graph, many members prefer not to fill out their skills, industries, etc. It's almost impossible to predict whether members' willing to work remotely if they have empty features. However, the job applications from members can also be considered as one type of ``member features''. In the other words, one can predict the skills and industries of members based on their job applications.
Moreover, the node features can be predicted based on all kinds of meta relations if heterogeneous graph is used. 
Here we define a \underline{F}eature \underline{R}estoration operator $\fr_w(\cdot)$ with trainable edge-wise weights $w^c_{ij}$. In the other words, each edge is assigned with a set of channel-specific scalar weights $\{w^c_{ij}\}_{c=1}^{N_c}$. $w^c_{ij}$ directly measures the similarity between node $v_i$ and $v_j$ in channel $c$ and doesn't contain any global information. This edge-wise treatment is especially suitable for the member-job application graph because of the relatively small neighborhood size. Many existing work used the attention mechanism~\cite{Velickovic2017,Wang2019,Kumar2019,Hu2020,Rossi2020,Zhu2021}, which contains graph-level vector $a$ and this weight is affected by the global information. This global vector suggests that attention weights are affected by all the nodes and edges in the whole graph. On the other hand, edge-wise weight $w^c_{ij}$ is not only channel-specific, but also only affected by two nodes $v_i$, $v_j$ and the edges $e_{ij}$ connecting them. Note that attention mechanism can be used before in the multi-channel learning in Section~\ref{sec:multi-channel}. Therefore, with the feature restoration implemented, our final model \proposedModel\ are able to capture both the global and local information.
To summerize, for each channel $c$, operator $\fr_w(\cdot)$ propagates the target node
features along the graph edges to get the restored source node features:
\begin{equation}
\bar{\mathbf{X}}^c=\fr_w{(\mathbf{X}^c,\ \mathcal{G})},
\end{equation}
and further compare with the original node features:
\begin{equation}
\begin{split}
\min_w\mathcal{L}&=\mathbb{E}_{\mathcal{U},\mathcal{V}}[\ell(\bar{\mathbf{X}}^c,\ \mathbf{X}^c)],\\
&=\mathbb{E}_{\mathcal{U},\mathcal{V}}[\ell(\fr_w(\mathbf{X}^c,\ \mathcal{G}),\ \mathbf{X}^c)]
\end{split}
\end{equation}

In particular, our feature propagation $\fr_w{(\mathbf{X}^c,\ \mathcal{G})}$ defines the direct edge-wise weights $w_{v\to u}$ (from job to member), $w_{u\to v}$ (from member to job) and only uses the first neighbors $\mathcal{N}$ for the computation effectiveness:
\begin{equation}
\fr_w{(\mathbf{X}^c,\ \mathcal{G})}=g_{w_{v\to u}}(\mathbf{X}^c,\ \mathcal{N})+g_{w_{u\to v}}(\mathbf{X}^c,\ \mathcal{N}),
\end{equation}
with
\begin{equation}
\begin{split}
\bar{x}^c_u&=g_{w_{v\to u}}(x^c_u,\ \mathcal{N}(u))=\sum_{v\in\mathcal{V}}w_{v\to u}e_{uv}x^c_v,\\
\bar{x}^c_v&=g_{w_{u\to v}}(x^c_v,\ \mathcal{N}(v))=\sum_{u\in\mathcal{U}}w_{u\to v}e_{uv}x^c_u,
\end{split}
\end{equation}
and $e_{uv}=1$ ($0$) denotes the (non-) existence of edge.

Then the loss function is written as 
\begin{equation}
\begin{split}
\min_{w}\mathcal{L}&=\min_{w_{v\to u},w_{u\to v}}\mathcal{L},\\
&=\mathbb{E}_{\mathcal{U},\mathcal{V}}[\mathcal{\ell}(g_{w_{v\to u}}(\mathbf{X}^c,\ \mathcal{N})+g_{w_{u\to v}}(\mathbf{X}^c,\ \mathcal{N}),\ \mathbf{X}^c),\\
&=\mathbb{E}_{\mathcal{U},\mathcal{V}}[\mathcal{\ell}((g_{w_{v\to u}}+g_{w_{u\to v}})(\mathbf{X}^c,\ \mathcal{N}),\ \mathbf{X}^c)].
\end{split}
\end{equation}

\begin{figure*}[htbp]
    \centering
    \includegraphics[width=0.9\linewidth]{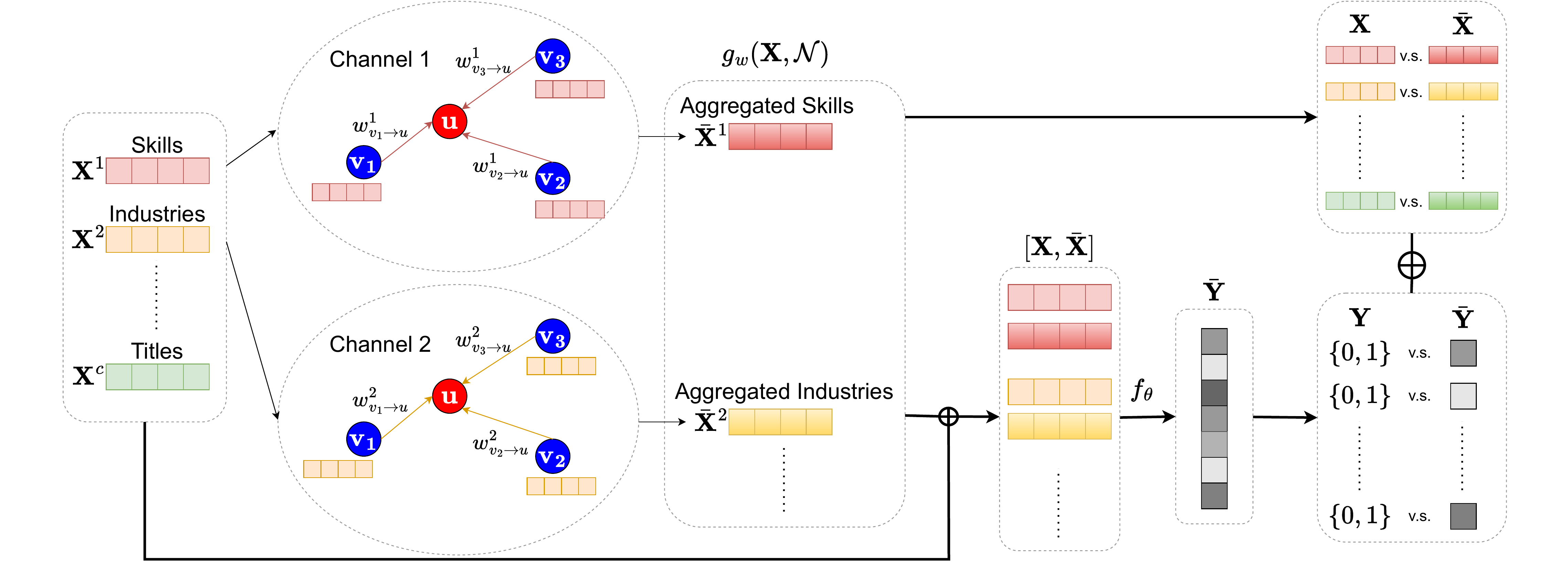}
    \caption{The flow of multi-object learning task equipped with multiple channels and feature restoration. The weights of the feature restoration operator and the parameters of neural network are trained jointly in the same backpropagation step.}
    \label{fig:feature-propagation}
\end{figure*}

\subsection{Multi-Object Learning}\label{sec:forward-backward}
After defining the multi-channel learning in \ref{sec:multi-channel} and feature restoration in \ref{sec:feature-restoration}, we are finally able to solve Problem \ref{p1}-\ref{p3}.

For forward computation, we first aggregate the member (job) features to job (member) features along the graph edges to recover the job (member) missing features $\bar{\mathbf{X}}^c$ for each channel $c$. Then the concatenation of the original node features and the restored node features $[\mathbf{X}^c,\ \bar{\mathbf{X}}^c]$ are fed into the multi-channel graph model to get the node embeddings $\mathbf{Z}^c$. Finally, the node-level remoteness prediction $\mathbf{Y}$ is obtained using the concatenated node embeddings $\mathbf{Z}$. More details are in Alg. \ref{alg:forward}.

For backward computation, multi-object learning is used to ensure the accuracy of both the restored information and remoteness prediction. The weights $w$ of the feature propagation and the parameters $\theta$ of the model are learned together using the following cost function:
\begin{equation}
\min_{w_{v\to u},w_{u\to v},\theta^c}\mathcal{L}(\beta_1,\beta_2)=\beta_1\min_{w_{v\to u},w_{u\to v}}\mathcal{L}_1+\beta_2\min_{\theta^c}\mathcal{L}_2,
\end{equation}
with
\begin{equation}
\begin{split}
\mathcal{L}_1&=\mathbb{E}_{c}[\mathbb{E}_{\mathcal{U},\mathcal{V}}[\mathcal{\ell}((g_{w_{v\to u}}+g_{w_{u\to v}})(\mathbf{X}^c,\ \mathcal{N}),\ \mathbf{X}^c)]],\\
\mathcal{L}_2&=\mathbb{E}_{\mathcal{U},\mathcal{V}}[\mathcal{\ell}(\dec(\concat(\\
&\hspace{2cm}\{f_{\theta^c}\circ(g_{w_{v\to u}}+g_{w_{u\to v}})(\mathbf{X}^c,\ \mathcal{G})\}_c,\ \mathbf{Y})].
\end{split}
\end{equation}

By default, we have $\beta_1=\beta_2=1.0$ without further notice.

\begin{algorithm}
\caption{Multi-Object Learning}
\begin{flushleft}
\textbf{Input:} Undirected graph $\mathcal{G}$ with members $\mathcal{U}$, jobs $\mathcal{V}$, edges $\mathcal{E}$;\\
Multi-channel node features $\mathbf{X}=\{\mathbf{X}_1,\cdots,\mathbf{X}_c\}$;\\
\textbf{Output:} Node embeddings $\mathbf{Z}$; Aggregated node features $\bar{\mathbf{X}}$; Node remoteness predictions $\bar{\mathbf{Y}}$;
The weights $w$ of Feature Restoration operator and parameters $\theta$ of content-based or graph-based model.\\
\end{flushleft}
\begin{algorithmic}[1]
\State Initial state $\mathbf{z}^{c,(0)}_i=\mathbf{x}^c_i$
\For{$c\in\{1,\cdots,N_c\}$}
    \State Compute the restored features
    
    \hspace{16pt}$\bar{\mathbf{X}}^c\gets g_{w_{v\to u}}(\mathbf{X}^c,\ \mathcal{N})+g_{w_{u\to v}}(\mathbf{X}^c,\ \mathcal{N})$
    \State Concatenation $\bar{\mathbf{X}}^c\gets \concat{(\mathbf{X}^c,\ \bar{\mathbf{X}}^c)}$
    \State Feature transformation $\bar{\mathbf{X}}^c\gets\linear^{N_{in}}_{N_{out}}(\bar{\mathbf{X}}^c)$
    \State Compute the channel-specific node embedding    
    
    \hspace{16pt}$\mathbf{Z}^c\gets f_{\theta^c}(\bar{\mathbf{X}}^c,\mathcal{G})$
\EndFor
\State Concatenate the learned embedding from all channels 

\hspace{16pt}$\mathbf{Z}\gets \concat{(\mathbf{Z}^1,\cdots,\mathbf{Z}^{N_c})}$

\State Hidden layer $\mathbf{H}=\sigma(\linear^{m}_d(\mathbf{Z}))$
\State Predict the workplace types 

\hspace{16pt}$\bar{\mathbf{Y}}=\sigmoid(\linear^{d}_1(\mathbf{H}))$

\State Loss function $\min_{w_{v\to u},w_{u\to v},\theta^c}\beta_1\mathcal{L}_1+\beta_2\mathcal{L}_2$
\State Back propagation and update parameters
\State \Return $\mathbf{Z},\bar{\mathbf{X}},\bar{\mathbf{Y}},w,\theta$
\end{algorithmic}
\label{alg:forward}
\end{algorithm}

\section{Experiments}\label{sec:experiment}
In this section, we will conduct experiments to validate the accuracy of the proposed methods. In particular, we will try to measure the following metrics:
\begin{itemize}
    \item The average precision of the remoteness predictions.
    \item The accuracy rate between the restored node features and the original node features.
\end{itemize}

and answer the following questions through conducting offline and online experiments using real-world data: 
\begin{itemize}
    \item How effective is the proposed algorithm comparing against existing approaches?
    \item How does each of the component of the proposed framework differently impact the performance of the model?
    \item Can the restored node features be used for other tasks, even the link-level/graph-level tasks?
\end{itemize}

\subsection{Data}
In all experiments, member-job application dataset is used. In this dataset, there exist $7106$ members, $42061$ jobs and $47990$ member-job edges. Originally, $2.5\%$ nodes have empty features. In order to further test the performance of multi-channel learning with feature propagation, the percentage of empty features is increased to $12.5\%/25\%$ in the second/third experiment. 

There are mainly three types of node features: a) \textbf{member and job titles} are scalars and are treated using embedding lookup table; b) \textbf{member and job skills} are one-hot vectors and represent the skills members have and jobs require. c) \textbf{member and job industries} are one-hot vectors and denote the industries of the members and jobs. Note that one member can be in several different industries. By default, the dimensions of skill and industry space are $3826$ and $151$ respectively. The whole space is used without further notice as in Section \ref{sec:performance}. In the later model analysis (Section \ref{sec:model-analysis}), we are going to decrease the dimension of skill space and focus on the most-used skill subspace, in order to see how the models behave with the limited node features.

\subsection{Experiment Setup}

All the models are trained on the $80\%$ data, validated on the $10\%$ data and tested on the $10\%$ data. The main tasks are the node-level classification problem, which classifies members/jobs as onsite or remote. The used metrics are average precision.

\paragraStartHighlight{Hyperparameters} For all the training, we use the Adam optimizer with a batch size in $\{1000,2000,4000,8000,10000\}$, a learning rate in $\{0.001,0.003,0.005,0.007,0.01\}$ and a dropout rate in $\{0.3,0.5\}$. The whole neighborhood is used since the average number of neighbors is around $10$ in each neighborhood. 

\paragraStartHighlight{Baselines} For all the baselines, we consider a few popular and representative state-of-the-art content-based and graph-based models, including: a) MLP, fully connected layers with RELU activation; b) GCN, Graph Convolutional Networks~\cite{Kipf2016}; c) GraphSAGE, with mean-based or LSTM-based aggregators~\cite{Hamilton2017}; d) GAT, Graph Attention Networks with $2$ multi-heads~\cite{Velickovic2017}.

\subsection{Performance Comparison with Baselines} \label{sec:performance}
\begin{table}[h!]
\begin{center}
\begin{tabular}{ |c|p{1.75cm}|p{1.75cm}|p{1.75cm}|p{1.75cm}|p{1.75cm}|p{1.75cm}|} 
\hline
\multicolumn{7}{|c|}{Member-Job Application Dataset} \\
\hline
Task & \multicolumn{6}{|c|}{Remoteness Prediction}  \\
\hline
Node Type &\multicolumn{1}{|c|}{Member}&\multicolumn{1}{|c|}{Job}&\multicolumn{1}{|c|}{Member}&\multicolumn{1}{|c|}{Job}&\multicolumn{1}{|c|}{Member}&\multicolumn{1}{|c|}{Job}\\
\hline
Missing Ratio &\multicolumn{2}{|c|}{$\sim2.5\%$}&\multicolumn{2}{|c|}{$\sim12.5\%$}&\multicolumn{2}{|c|}{$\sim25\%$}\\
\hline
MLP                     & $80.16\pm0.2$ & $68.04\pm0.5$& $79.89\pm1.4$ & $62.20\pm0.5$& $79.54\pm1.0^{\ddagger}$ & $59.20\pm0.2$  \\
GCN                     & $60.85\pm0.7$ & $66.02\pm0.5$& $60.17\pm1.0$ & $64.57\pm0.1$& $57.08\pm2.4$ & $63.83\pm0.1$  \\
GAT                     & $72.05\pm0.5$ & $66.96\pm0.5$& $67.38\pm0.1$ & $66.15\pm0.4$& $63.34\pm0.9$ & $65.31\pm0.4$ \\
GraphSAGE               & $82.18\pm1.1^{\ddagger}$ & $69.06\pm0.5^{\ddagger}$& $80.36\pm0.4^{\ddagger}$ & $67.99\pm0.2^{\ddagger}$& $76.40\pm0.5$ & $66.97\pm0.9^{\ddagger}$ \\
\proposedModel  & $84.91\pm0.3^{\dagger}$ & $69.57\pm0.1^{\dagger}$& $85.03\pm0.6^{\dagger}$ & $69.37\pm0.6^{\dagger}$& $84.71\pm0.5^{\dagger}$ & $68.55\pm0.7^{\dagger}$ \\
\hline
\end{tabular}
\end{center}
\caption{Accuracy \% for remoteness prediction task. \textbf{First}$^{\dagger}$, \textbf{Second}$^{\ddagger}$ best performing method. All the results are averaged over $3$ runs on the same training/validation/testing dataset.}
\label{tab:member-job}
\end{table}

In Table \ref{tab:member-job}, \proposedModel\ outperforms the traditional MLP and Graph models on both the members and jobs for all the experiments, especially when the percentage of the missing features is larger. When there are $2.5\%$ data missing, \proposedModel\ outperforms the GraphSAGE by $\sim2.8\%$ on members and $\sim0.5\%$ on jobs. With $25\%$ data missing, \proposedModel\ outperforms the second best models by $\sim5.2\%$ on members and $\sim1.6\%$ on jobs. This improvement will become larger with more data missing, as in Section \ref{sec:model-analysis}. 

Moreover, \proposedModel\ can also predict the node missing features while all the other models are unable to do. These restored features from the trained models can be used later for other tasks to help improve the corresponding performance.

\subsection{Model Analysis} \label{sec:model-analysis}
\begin{figure}[H]
\centering
\begin{subfigure}[c]{0.45\linewidth}
  \includegraphics[width=0.9\linewidth]{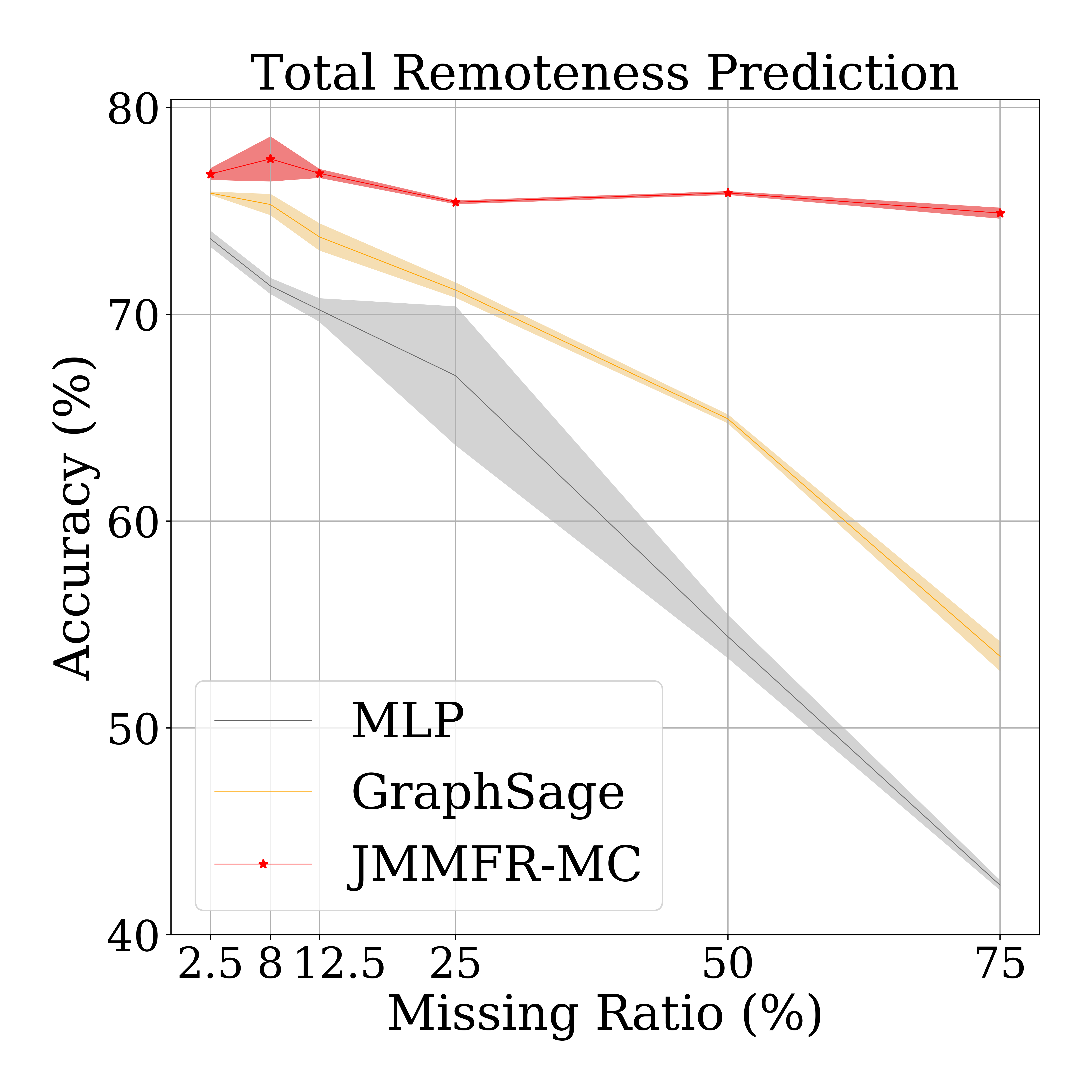}
\end{subfigure}
\begin{subfigure}[c]{0.45\linewidth}
  \includegraphics[width=0.9\linewidth]{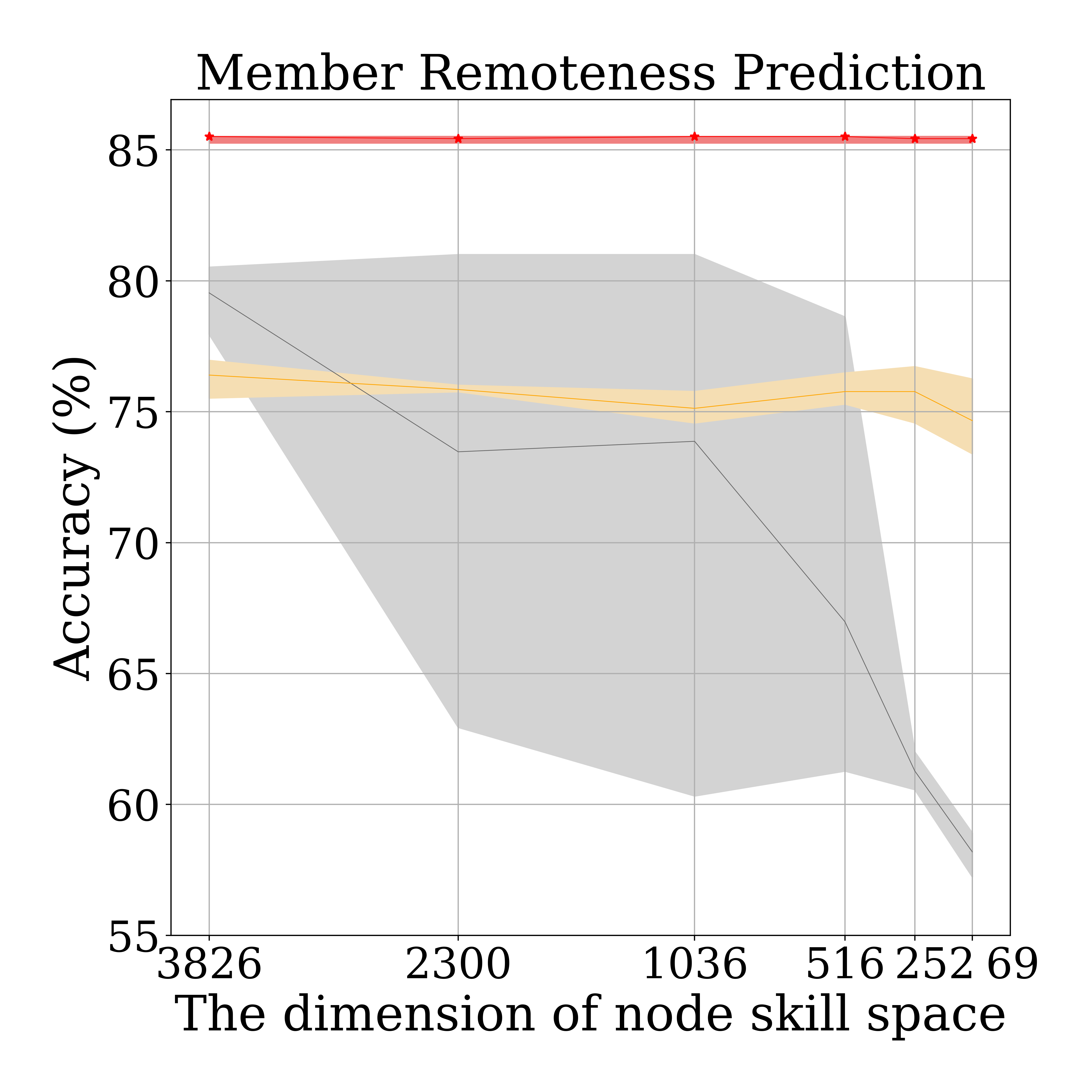}
\end{subfigure}
\hfill
\begin{subfigure}[c]{0.45\linewidth}
    \caption{Missing Ratio}
    \label{fig:node_feature}
\end{subfigure}
\begin{subfigure}[c]{0.45\linewidth}
    \caption{Limited Feature Space}
    \label{fig:node_skill}
\end{subfigure}
\caption{(\subref{fig:node_feature}) \proposedModel\ is robust to the ratio of the missing features. We increased the ratio of missing data from original $2.5\%$ to $75\%$ and tested the performance of MLP/GraphSAGE/\proposedModel. (\subref{fig:node_skill}) \proposedModel\ is robust to the dimension of the node feature space. We decreased the number of node features from a total of $3826$ skills to $69$ most-used skills and tested the performance of MLP/GraphSAGE/\proposedModel.}
\end{figure}

\subsubsection{Robustness against different data missing ratios}
In this experiment, we verify the robustness of \proposedModel\ against the nodes with missing features by varying the missing ratios and comparing the results with other models. The missing ratios are in $\{2.5\%,8.0\%,12.5\%,25.0\%,50.0\%,75.0\%\}$.

As the missing ratio increases, the total accuracy of MLP and GraphSAGE ``linearly'' decreases. It makes sense since that MLP and GraphSAGE tend to give wrong predictions on the nodes with missing features. On the other hand, the performance of \proposedModel\ stays stable since \proposedModel\ can recover the node missing features using the feature restoration operator.

\subsubsection{Robustness against the limited feature space}
In this experiment, we verify the robustness of\proposedModel\ against the nodes with limited feature space by shrinking the feature space and comparing the results with other models. We ranked all the member and job skills according to coverage on the data, where skills with least null values will be ranked top. Then the top $D_{\text{skill}}$ most-owned skills are chosen to be the new skill subspace $\{0,1\}^{D_{\text{skill}}}$ and $D_{\text{skill}}\in\{3826,2300,1036,516,252,69\}$. As the dimension of subspace shrinks, the accuracy of MLP drops, while GraphSAGE and \proposedModel\ are relatively stable, especially \proposedModel. It shows the ability of \proposedModel\ to learn on limited feature space.

\begin{table}[h!]
\begin{center}
\resizebox{3.0in}{!}{%
\begin{tabular}{ |c|p{1.75cm}|p{1.75cm}|} 
\hline
\multicolumn{3}{|c|}{Member-Job Application Dataset} \\
\hline
Task & \multicolumn{2}{|c|}{Aggregated Feature} \\
\hline
Data Missing Ratio & \multicolumn{1}{|c|}{Skills}&\multicolumn{1}{|c|}{Industries}\\
\hline
$\sim\ \ 2.5\%$      &$65.18\pm0.2$&$86.71\pm0.2$\\
\hline
$\sim12.5\%$         &$65.62\pm0.1$&$86.70\pm0.1$\\
\hline
$\sim25.0\%$         &$65.11\pm0.1$&$86.56\pm0.1$\\
\hline
$\sim50.0\%$         &$63.59\pm0.1$&$85.61\pm0.1$\\
\hline
$\sim75.0\%$         &$57.09\pm0.2$&$81.17\pm0.3$\\
\hline
\end{tabular}}
\end{center}
\caption{Average precision \% for aggregated feature prediction. All the results are averaged over $3$ runs on the same training/validation/testing dataset.}
\label{tab:aggregated-feature}
\end{table}

\subsubsection{Generalization of the restored features}

Here we discuss the possibility in the generalization of the restored features. As in Table~\ref{tab:aggregated-feature}, our restored features maintain the accuracy for all missing levels, which implies the possibility to use the restored features further for other link-level and graph-level tasks, such as job recommendation. We are currently working in this direction.

\begin{figure}[H]
\centering
\begin{subfigure}[c]{.24\textwidth}
    \includegraphics[width=\textwidth]{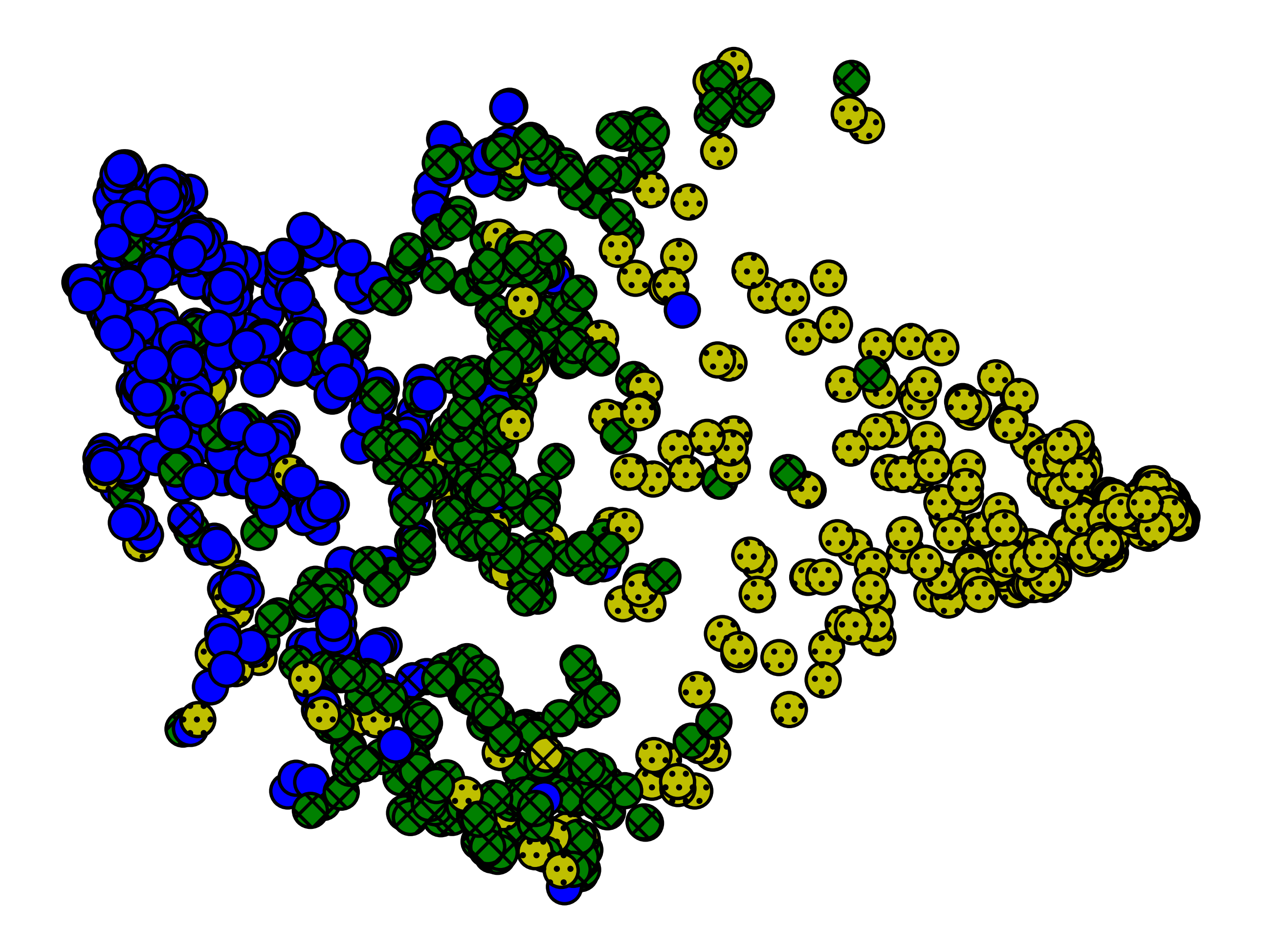} 
\end{subfigure}
\hfill
\begin{subfigure}[c]{.24\textwidth}
    \includegraphics[width=\textwidth]{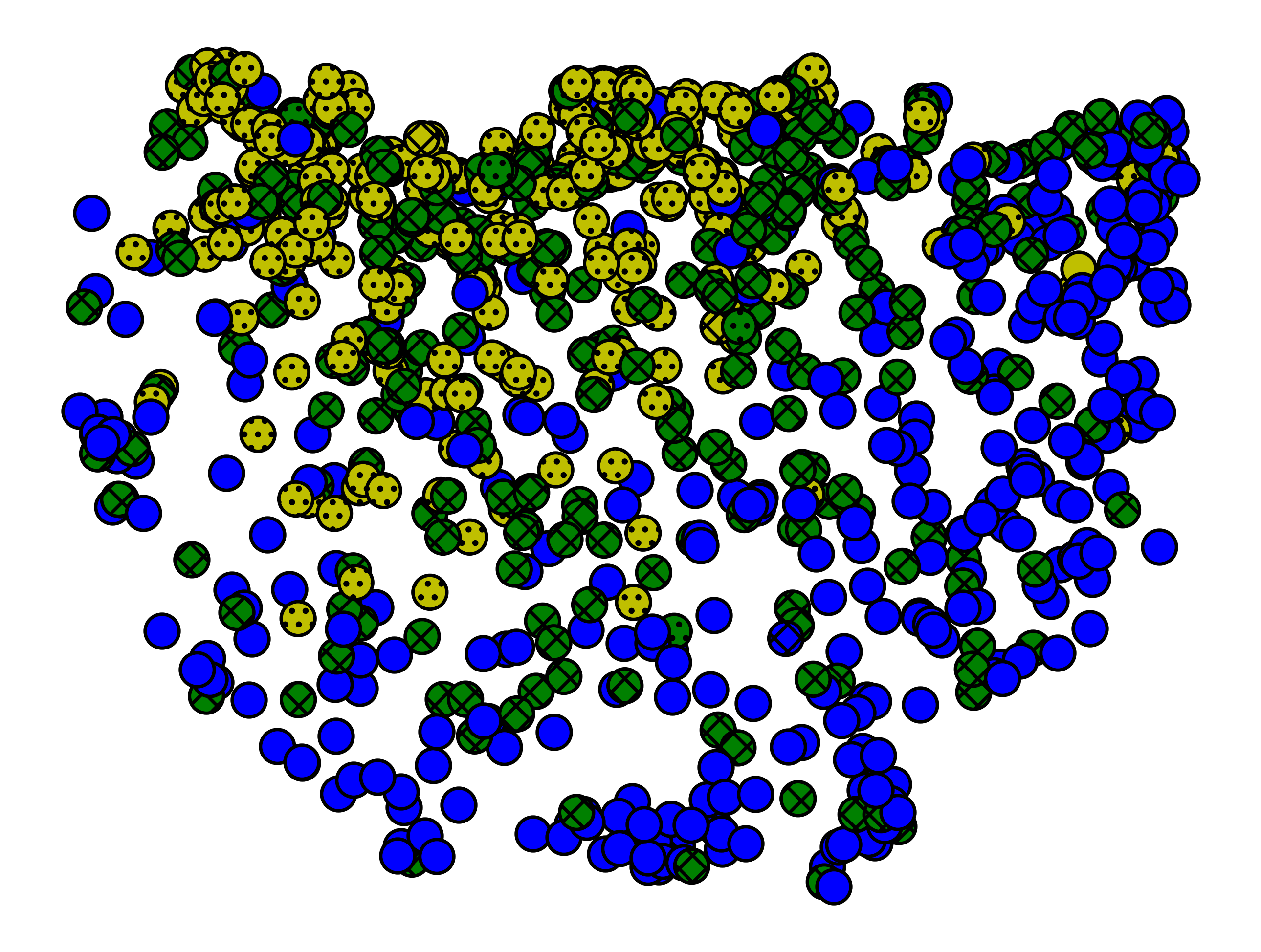} 
\end{subfigure}
\hfill
\begin{subfigure}[c]{.24\textwidth}
    \includegraphics[width=\textwidth]{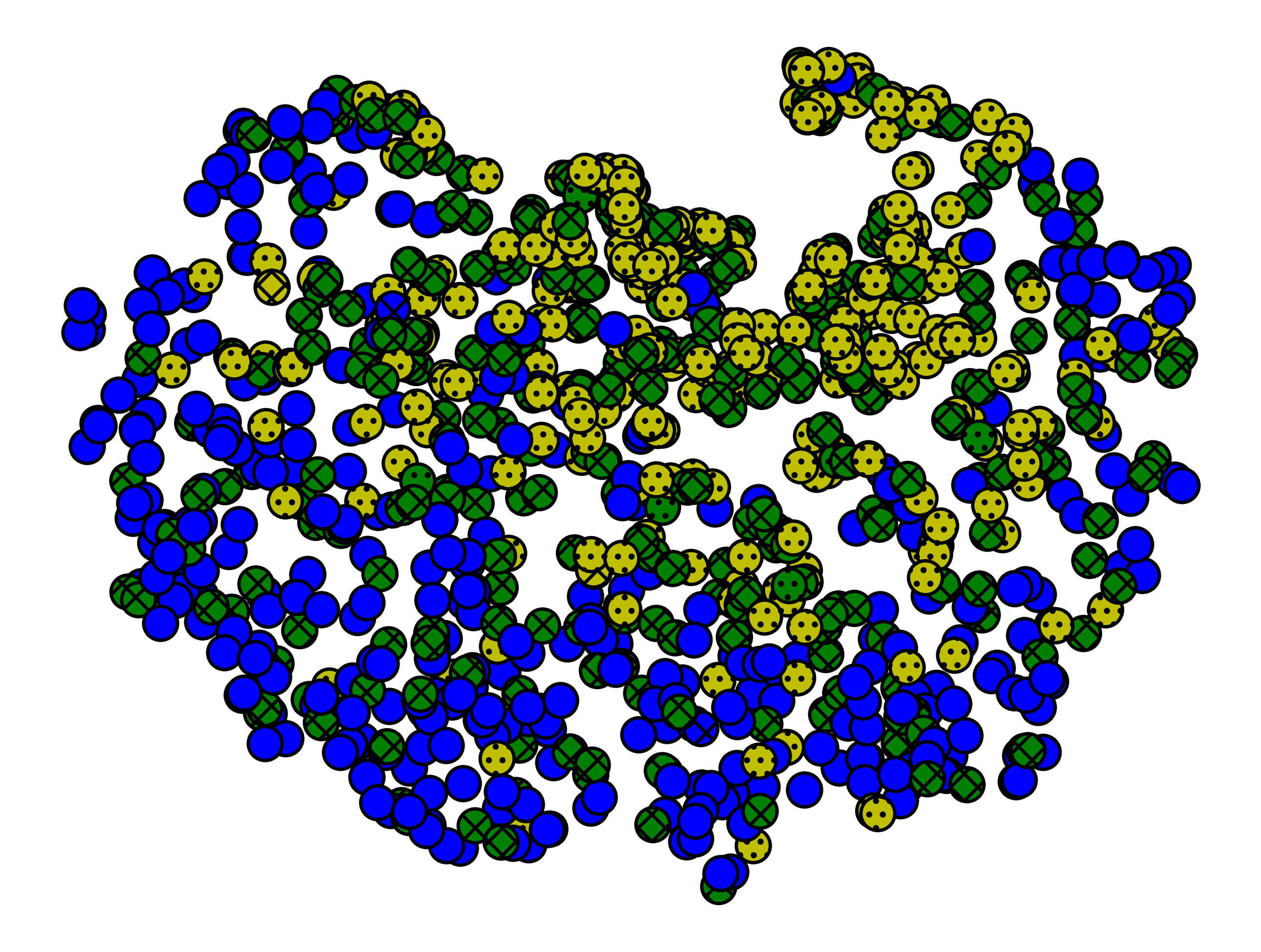} 
\end{subfigure}
\hfill
\begin{subfigure}[c]{.24\textwidth}
    \includegraphics[width=\textwidth]{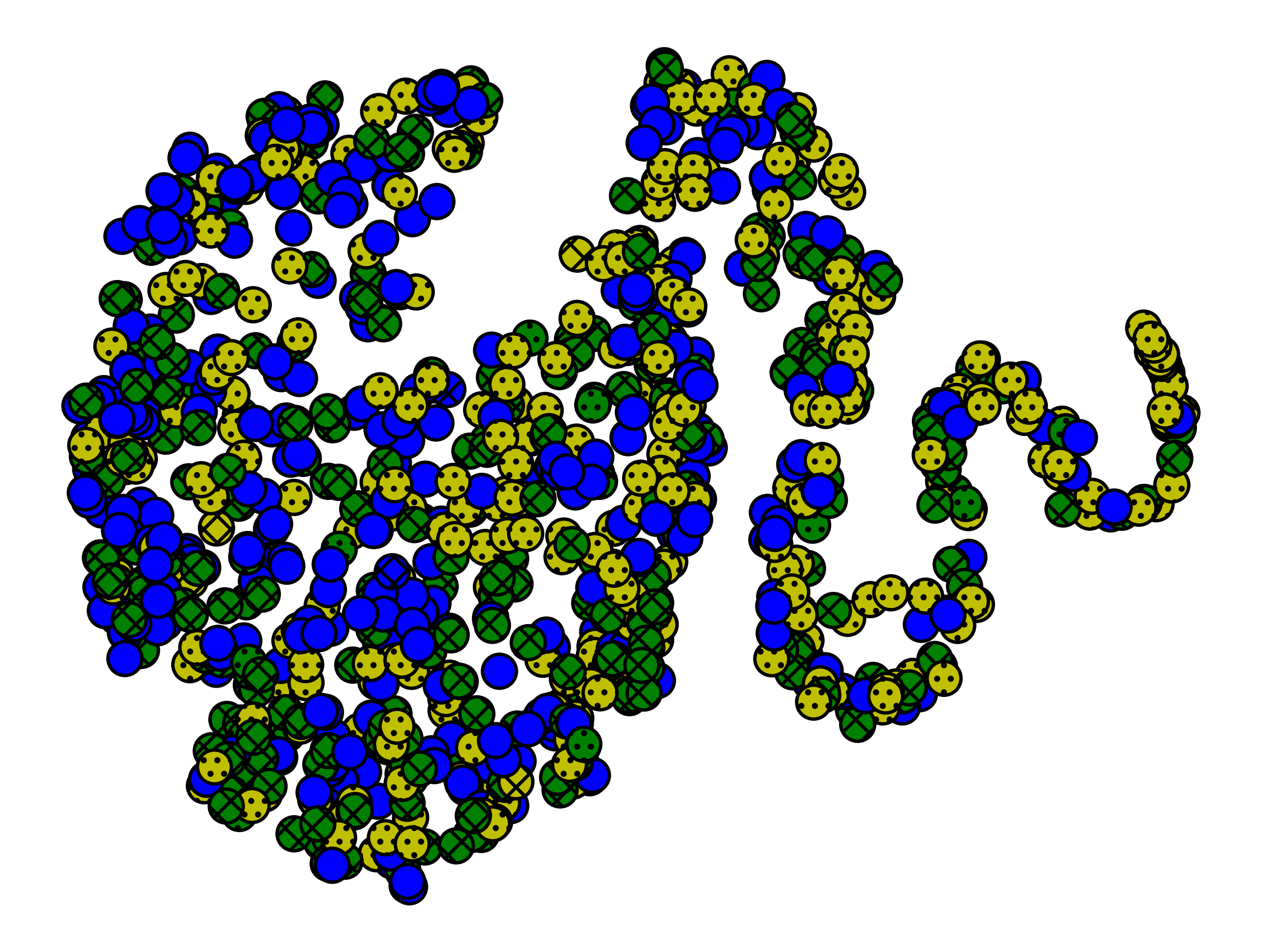} 
\end{subfigure}
\begin{subfigure}[c]{.24\textwidth}
    \caption{\proposedModel}
    \label{fig:multisage_skill_space}
\end{subfigure}
\hfill
\begin{subfigure}[c]{.24\textwidth}
    \caption{GAT}
    \label{fig:gat_skill_space}
\end{subfigure}
\hfill
\begin{subfigure}[c]{.24\textwidth}
    \caption{GraphSAGE}
    \label{fig:graphsage_skill_space}
\end{subfigure}
\hfill
\begin{subfigure}[c]{.24\textwidth}
    \caption{GCN}
    \label{fig:gcn_skill_space}
\end{subfigure}
\caption{Visualization of two-dimensional t-SNE transformed node embeddings from GNN models. Nodes with the same color own the same skill. For nodes with more than one skill, random generator is used to randomly pick one of skills. (\subref{fig:multisage_skill_space}) Nodes with different skills are well separated in\ \proposedModel. (\subref{fig:gat_skill_space})(\subref{fig:graphsage_skill_space})(\subref{fig:gcn_skill_space}) Other baselines fail to separate the embedding space to some extent.}
\label{fig:t-sne}
\end{figure}

\subsubsection{Visualization of Embedding space}


Finally, we validate the effectiveness of \proposedModel qualitatively by providing the visualization of the t-SNE transformed
feature embeddings on \proposedModel\ and other baseline graph models, as in Figure~\ref{fig:t-sne}. Note that
these node embeddings correspond to the three different skills of the dataset, verifying the model’s discriminative power across the node skills in the real-world dataset.

In order to do this, we first investigate the node embeddings for various skills. For each skill, we take the average of all member embeddings with this skill. Note that members have multiple skills, so the same member embeddings are used multiple times for multiple skills. 
After that, the resulting midpoints of different skills are separated from each other and we pick the three most-seperated skills in order to visualize the node embeddings with different skills.

After picking these three skill points for each graph models, we can use t-SNE to visualize the node embeddings in Figure~\ref{fig:t-sne}. From the visualizations, we can see that node embeddings are well separated for \proposedModel\ as in Figure~\ref{fig:multisage_skill_space}, while the node embeddings are less separated and tend to be mixed with each other for other graph models, as in Figure~\ref{fig:gat_skill_space},\ref{fig:graphsage_skill_space},\ref{fig:gcn_skill_space}.

\section{Related Work}\label{sec:related-work}

\paragraStartHighlight{Graph Models}, such as GCN~\cite{Kipf2016}, GraphSAGE~\cite{Hamilton2017} and GAT~\cite{Velickovic2017}, didn't consider the missing feature problem and rich semantic information in the member-job. Besides that, this work is the first time that graph models are leveraged in the remoteness problem.

\paragraStartHighlight{Models on Missing Feature Problem}, such as SAT~\cite{Chen2022}, PaGNN~\cite{Jiang2020}, GCNMF~\cite{Taguchi2021} also didn't tackle the missing feature problem and utilize the multi-channel learning. Moreover, their scalability is limited for the large real-world online dataset since the adjacency matrix is used in their implementation.

\section{Conclusion and Future Work}\label{sec:con}

In this work, we aim to close the gap in the remote workplace prediction task and tackle the missing feature problem jointly. Missing feature problem can be considered as a cold-state problem and is common in the real-world job recommender platform because of the limited node and edge information for less active job seekers. We observe that for a considerable large proportion of members and jobs, node features are missing or less adequate, while traditional models normally assume a complete set of features. To tackle this problem, we propose a job-member embedding model with feature restoration through multi-channel learning (JMMFR-MC), which outperforms the state-of-the-art graph models for all missing ratios. In \proposedModel, we introduce a
novel way to restore the node missing features using the graph structure and multi-channel learning that utilizes both the node-level and channel-level weights. Experimental results on real-world datasets show that \proposedModel\ outperforms the state-of-the-art baselines on the average precision. Besides that, \proposedModel\ is robust and qualitatively have better discriminative power. Future work can be focused on two directions. First, we would like to investigate if restored features can be used on other tasks, such as link-level and graph-level tasks. Second, more edge types can be included, such as member \emph{skip} job and member \emph{save} job, and heterogeneous graph models can be discussed.

\balance
\balance

\bibliographystyle{unsrt}  
\bibliography{reference}

\end{document}